\documentclass[aps,prl,twocolumn,superscriptaddress]{revtex4}%
\usepackage{epsfig,dsfont,amssymb,amsmath,amsthm,amsfonts,amsbsy,mathrsfs}
\usepackage{graphicx}
\usepackage{amsmath}
\usepackage{amssymb}
\usepackage[utf8x]{inputenc}
\usepackage{graphicx}
\usepackage{dcolumn}
\usepackage{bm}
\usepackage{xcolor}
\usepackage{amsmath}
\usepackage{chngcntr}
\usepackage{nicefrac}
\usepackage[colorlinks,linkcolor=blue,hyperindex,CJKbookmarks]{hyperref}
\begin{document}

\title{
Coexistence of Multiple Stacking Charge Density Waves in Kagome Superconductor ${\mathrm{CsV}}_3{\mathrm{Sb}}_5$}

\author{Qian Xiao}\thanks{These authors contributed equally to this work.}
\affiliation{International Center for Quantum Materials, School of Physics, Peking University, Beijing 100871, China}

\author{Yihao Lin}\thanks{These authors contributed equally to this work.}
\affiliation{International Center for Quantum Materials, School of Physics, Peking University, Beijing 100871, China}

\author{Qizhi Li}\thanks{These authors contributed equally to this work.}
\affiliation{International Center for Quantum Materials, School of Physics, Peking University, Beijing 100871, China}

\author{Xiquan Zheng}\thanks{These authors contributed equally to this work.}
\affiliation{International Center for Quantum Materials, School of Physics, Peking University, Beijing 100871, China}

\author{Sonia Francoual}
\affiliation{Deutsches Elektronen-Synchrotron (DESY), Hamburg D-22607, Germany}

\author{Christian Plueckthun}
\affiliation{Deutsches Elektronen-Synchrotron (DESY), Hamburg D-22607, Germany}

\author{Wei Xia}
\affiliation{School of Physical Science and Technology, ShanghaiTech University, Shanghai 201210, China}
\affiliation{ShanghaiTech Laboratory for Topological Physics, ShanghaiTech University, Shanghai 201210, China}

\author{Qingzheng Qiu}
\affiliation{International Center for Quantum Materials, School of Physics, Peking University, Beijing 100871, China}

\author{Shilong Zhang}
\affiliation{International Center for Quantum Materials, School of Physics, Peking University, Beijing 100871, China}

\author{Yanfeng Guo}
\affiliation{School of Physical Science and Technology, ShanghaiTech University, Shanghai 201210, China}
\affiliation{ShanghaiTech Laboratory for Topological Physics, ShanghaiTech University, Shanghai 201210, China}

\author{Ji Feng}
\email{jfeng11@pku.edu.cn}
\affiliation{International Center for Quantum Materials, School of Physics, Peking University, Beijing 100871, China}
\affiliation{CAS Center for Excellence in Topological Quantum Computation, University of Chinese Academy of Sciences, Beijing 100190, China}

\author{Yingying Peng}
\email{yingying.peng@pku.edu.cn}
\affiliation{International Center for Quantum Materials, School of Physics, Peking University, Beijing 100871, China}
\affiliation{Collaborative Innovation Center of Quantum Matter, Beijing 100871, China}

\date{\today}

\begin{abstract}
The recently discovered Kagome family ${\mathrm{AV}}_3{\mathrm{Sb}}_5$ (A = K, Rb, Cs) exhibits rich physical phenomena, including non-trivial topological electronic structure, giant anomalous Hall effect, charge density waves (CDW) and superconductivity. Notably, CDW in ${\mathrm{AV}}_3{\mathrm{Sb}}_5$ is evidenced to intertwine with its superconductivity and topology, but its nature remains elusive. Here, we combine x-ray scattering experiments and density-functional theory calculations to investigate the CDWs in ${\mathrm{CsV}}_3{\mathrm{Sb}}_5$ and demonstrate the coexistence of 2 $\times$ 2 $\times$ 2 and 2 $\times$ 2 $\times$ 4 CDW stacking phases. 
Competition between these CDW phases is revealed by tracking the temperature evolution of CDW intensities, which also manifests in different transition temperatures during warming- and cooling measurements.  
We also identify a meta-stable quenched state of ${\mathrm{CsV}}_3{\mathrm{Sb}}_5$ after fast-cooling process. Our study demonstrates the coexistence of competing CDW stackings 
in ${\mathrm{CsV}}_3{\mathrm{Sb}}_5$, offering new insights in understanding the novel properties of this system.
\end{abstract}

\maketitle

Materials with kagome nets are promising candidates for quantum spin liquid \cite {S.Yan2011science,2016RevModPhys,2017RevModPhys} and can support a variety of electronic orders like charge bond order, charge density wave (CDW) and unconventional superconductivity\cite{Guo.H.M.2009prb,2009prb.sc,2013prb.competing.Orders}.
Recently, a new class of kagome materials ${\mathrm{AV}}_3{\mathrm{Sb}}_5$ (A = K, Rb, Cs) has entered the scene and triggered a surge of research. These materials are found to be $Z_2$ topological metals and show giant anomalous Hall conductivity\cite{prm2019Brenden,Brenden.prl,AHE.KVS,AHE.CVS}. More intriguingly, superconductivity transitions below $\sim$ 3 K were reported in ${\mathrm{AV}}_3{\mathrm{Sb}}_5$ \,\cite{Brenden.prl,prm2021sc.KVS,cpl2021RVS.sc}. Above the superconducting transition temperature, a CDW phase transition ranging from 78 K to 103 K was also reported\,\cite{prm2019Brenden,Brenden.prl,cpl2021RVS.sc,jiangyx}.  
As in most unconventional superconductors, such as cuprates\,\cite{YBCO2012SCI,YBCO2012natphy} and pnictides\,\cite{BNCA2019PRL}, the CDW in ${\mathrm{AV}}_3{\mathrm{Sb}}_5$ seems to compete with its superconductivity \,\cite{song2021CVS.thickness,Qi2021pressure,uniaxial2021prb,RVS.pressure,KVS.pressure}.
In addition, CDW in kagome lattice has been proposed to be a chiral flux phase, which breaks the time-reversal symmetry (TRS) without spin polarization and may be related to the anomalous Hall effect\,\cite{zhiweiWang.prb,jiangyx,AHE.CVS,AHE.KVS,yu2021flux,xilinFeng2021}. This TRS breaking has been found to persist into the superconductivity state, indicating an intimate relation between charge order and unconventional superconductivity\,\cite{miuSR.nat}. Additionally, the nematic susceptibility is observed to be enhanced below CDW transition temperature by elastoresistance measurements\,\cite{nematic.CDW.nat}. Overall, these results indicate that the CDW in ${\mathrm{AV}}_3{\mathrm{Sb}}_5$ is intertwined with its superconductivity and topology in an intriguing way. 

Of the ${\mathrm{AV}}_3{\mathrm{Sb}}_5$ family, ${\mathrm{CsV}}_3{\mathrm{Sb}}_5$ has the highest superconducting transition temperature $T_{\mathrm{c}}$ $\sim$ 2.5 K\,\cite{Brenden.prl}. It also undergoes a CDW transition at $\sim$ 94 K with both in-plane and out-of-plane components, whose underlying mechanism remains elusive\,\cite{Brenden.prl,H.X.Li.IXS.prx,sdh2021Brenden}.
More strikingly, the superconducting phase diagram of ${\mathrm{CsV}}_3{\mathrm{Sb}}_5$ shows a double-dome while the CDW phase is diminished by applying pressure or doping, indicating an unusual competition between CDW and superconductivity \,\cite{F.H.Yu.pressure.nc,XuChen.pressure.cpl,ChenKY.pressure.PRL,doping.2sc}.
Regarding the in-plane CDW, while the inverse star of David (ISD) deformation is energetically favored according to first principles calculations and as confirmed by the observation of the Shubnikov-de
Haas oscillation\,\cite{H.X.Tan.prl.calcu,xilinFeng2021,sdH.prl.CVS}, the star of David (SD) deformation is supported by nuclear quadrupole resonance measurements\,\cite{J.luo2021NMR}.
As for the out-of-plane component, in addition to the proposed 
2 $\times$ 2 $\times$ 2 (2c) and 1 $\times$ 4 modulations
\,\cite{H.X.Li.IXS.prx,zhiweiWang.prb,zuoweiLiang.prx,jiangyx,shumiya.prb.STM,zhao.nat}, non-resonant x-ray diffraction measurements on ${\mathrm{CsV}}_3{\mathrm{Sb}}_5$ observed a unique 2 $\times$ 2 $\times$ 4 (4c) CDW phase appearing below 94 K\,\cite{sdh2021Brenden}, which has not been reported in ${\mathrm{KV}}_3{\mathrm{Sb}}_5$ or ${\mathrm{RbV}}_3{\mathrm{Sb}}_5$.
It is argued that there is only 4c CDW phase in ${\mathrm{CsV}}_3{\mathrm{Sb}}_5$\,\cite{sdh2021Brenden}, which needs further experimental confirmation.

The 2c and/or 4c CDW phases may originate from different inter-layer stackings. The stacking degree of freedom in layered materials has been shown to be a powerful knob for manipulating physical properties. For example, in bilayer graphene, flat bands and superconductivity are induced by twisting two layers of non-superconducting graphene by a ``magic angle''\,\cite{cao2018nat}. In transition metal dichalcogenides, CDW stacking is considered to be important in determining the metal-insulator transition\,\cite{CDWstacking2000TaS2,TaS22016natcom}. Therefore, it is important to investigate in detail the CDW phase(s) in ${\mathrm{CsV}}_3{\mathrm{Sb}}_5$, in order to understand the role it plays in various electronic properties in this material. 

\begin{figure}[htbp]
\centering
\includegraphics[width=0.5\textwidth]{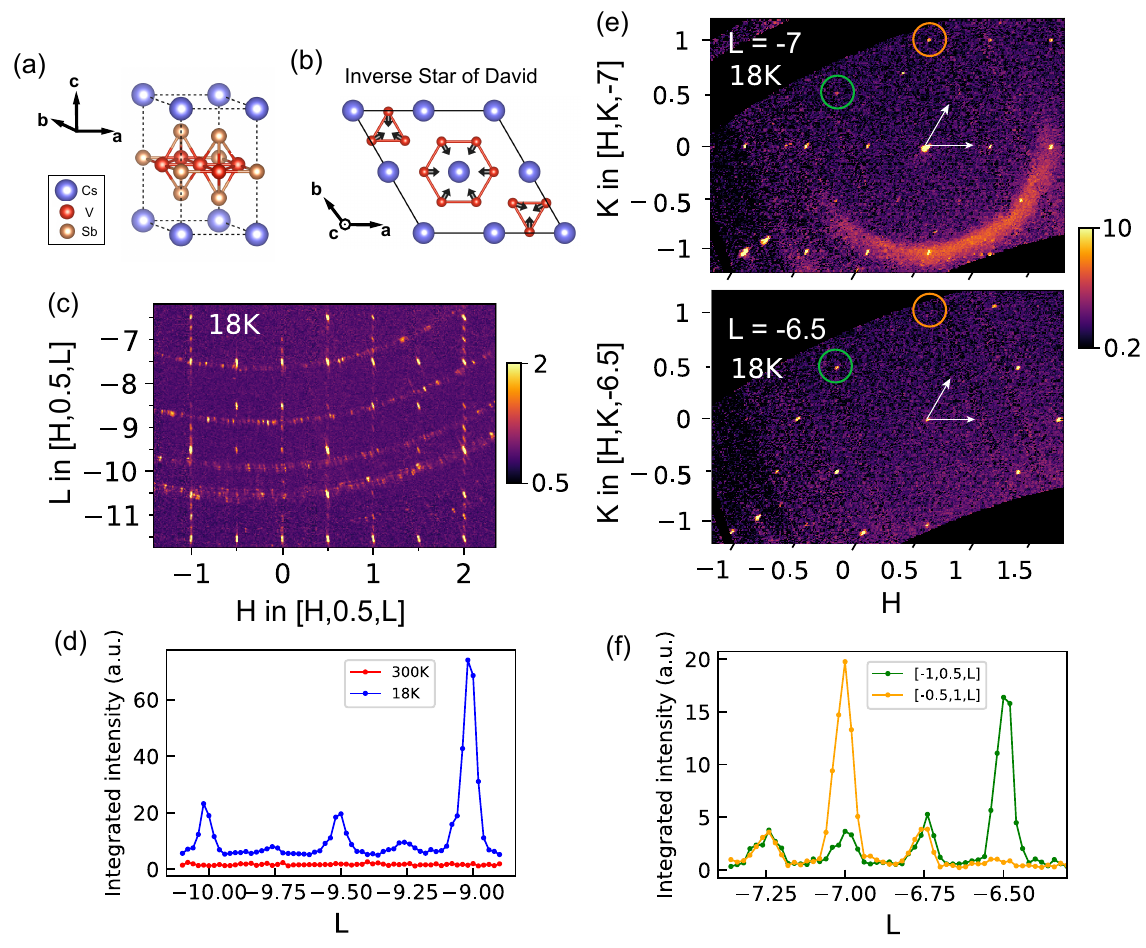}
\caption{(a) Crystal structure of ${\mathrm{CsV}}_3{\mathrm{Sb}}_5$ in normal state\,\cite{VESTA}. It belongs to P6/mmm space group (No.\,191)\,\cite{prm2019Brenden}. (b) Schematic plot of inverse Star of David (ISD). (c) $(H, L)$ map of reciprocal space at $K$ = 0.5 measured at 18K. The momentum axes are plotted in reciprocal length, i.e., $\AA^{-1}$. (d) $L$-cuts of [2.5, -3.5, L] at 300 K and 18 K, respectively. Offset is added for clarity. (e) $(H, K)$ maps of reciprocal space at $L$ = -7 (upper panel) and $L$ = -6.5 (lower panel) measured at 18K. The CDW peaks highlighted by green and orange circles are related to $C_6$ symmetry. Their cuts along the $L$-direction are shown in (f). In (c)(e),  
the arc signals come from beryllium domes, and the intensity of data is in logscale.
\label{fig1}}
\end{figure}

In this work, we use x-ray diffraction (XRD) and resonant x-ray scattering (REXS) techniques in conjunction with density-functional theory (DFT) calculations to study the CDW phases in ${\mathrm{CsV}}_3{\mathrm{Sb}}_5$. Based on the temperature evolution of diffraction peaks, we establish the coexistence of 2c and 4c CDW phases, with different transition temperatures. The observed spectral weight transfer between 2c and 4c diffraction peaks in warming and cooling processes signals a competition between these phases. In particular, the onset temperature of 4c on cooling ($\sim$ 89 K) is found to be lower than the corresponding vanishing temperature on warming ($\sim$ 92.8 K). An intriguing quenched state due to fast cooling is observed in the warming experiment, which could have critical implications in interpreting experimental anomalies at about 60 $\sim$ 70 K\,\cite{2symmetry2021natcom,NLWang2021prb,Qstahl2021}. By performing spatial mapping of CDW intensity using REXS, we observe that the different CDW modulations tend to exclude each other spatially further supporting the coexistence of different order parameters and competition between them. Comparison with computed diffraction patterns based on DFT calculations suggests the observed CDW structures arise from different stackings of ISD deformed V-containing layers. Our new results clarify the stacking CDWs in ${\mathrm{CsV}}_3{\mathrm{Sb}}_5$, which is very important in understanding and controlling the topological and superconducting properties in this material.

Single crystal x-ray diffraction measurements were performed using the custom-designed x-ray instrument equipped with a Xenocs Genix3D Mo $K_\alpha$ (17.48 keV) x-ray source, which provides $\simeq$ 2.5$\times 10^7$ photons/sec in a beam spot size of 150 $\mu m$ at the sample position\,\cite{SI}. Resonant x-ray scattering measurements were carried out at the beamline P09 at PETRA III, DESY (Hamburg, Germany)\,\cite{P09}. The single-crystalline ${\mathrm{CsV}}_3{\mathrm{Sb}}_5$ samples for the present study were grown by a self-flux growth method\,\cite{SI}.

The structure of ${\mathrm{CsV}}_3{\mathrm{Sb}}_5$ in normal state\,\cite{VESTA} is shown in FIG.~\ref{fig1}(a), which is comprised of alternating layers of alkali-metal and ${\mathrm{V}}_3{\mathrm{Sb}}_5$. The vanadium atoms form a perfect kagome lattice under ambient conditions\,\cite{prm2019Brenden}. After cooling from 300 K to 18 K (at a rate of 8 K/min), a cascade of CDW peaks becomes manifest at the half-integer $K$ plane, as shown in Figure~\ref{fig1}(c-d). Diffraction peaks are labeled by their Miller indices, $(H,K,L)$, of the undistorted high-temperature phase. The CDW peaks are about 3 orders of magnitude weaker than the main Bragg peaks, indicating small lattice distortions. 
The x-ray diffraction pattern was indexed according to a hexagonal unit cell with lattice parameters $a = b \simeq$ 5.53\,\AA \,and $c \simeq$ 9.28\,\AA~ \,\cite{SI}. 
The observed CDW peaks can be classified into three categories, namely, {\bf q}$_1$, {\bf q}$_2$ and {\bf q}$_3$-type  peaks corresponding to integer, half-integer and quarter-integer $L$ values, respectively. For a CDW reflection, either $H$ or $K$ or both are half-integers owing to the in-plane $2\times 2$ reconstruction\,\cite{jiangyx,zhiweiWang.prb}.

It is worth mentioning that we have not detected any reflection with a quarter integer $H$ or $K$, suggesting the absence of cell quadrupling along $a$ or $b$, which is consistent with previous XRD measurements\,\cite{li2021spatial}.
Thus, our results obtained with a highly sensitive two-dimensional detector further confirm that the 1 $\times$ 4 modulation observed by several STM experiments may not be a bulk effect\,\cite{li2021spatial,zhiweiWang.prb,zhao.nat}. 
We also notice that $C_6$ symmetry is broken at low temperatures, consistent with previous works\,\cite{Qstahl2021}. Because the CDW peaks at {\bf q} = [$-1,0.5,L$] and [$-0.5,1,L$] are equivalent in $C_6$ symmetry, however, as shown in Figure~\ref{fig1}(e, f), they differ greatly in intensity at 18 K. Either ISD [shown in Figure~\ref{fig1}(b)], SD distortion or order of orbital current would preserve $C_6$ symmetry in kagome plane. However, for a 3D CDW, the phase shift of adjacent kagome layers breaks the $C_6$ symmetry\,\cite{xilinFeng2021,miao2021geometry,H.X.Tan.prl.calcu}. One recent measurement of polar Kerr rotation of ${\mathrm{CsV}}_3{\mathrm{Sb}}_5$ has reported the formation of two-fold rotation symmetry below $T_{\mathrm{CDW}} \sim$ 94 K\,\cite{NLWang.TRS}.

\begin{figure}[htbp]
\centering
\includegraphics[width=0.5\textwidth]{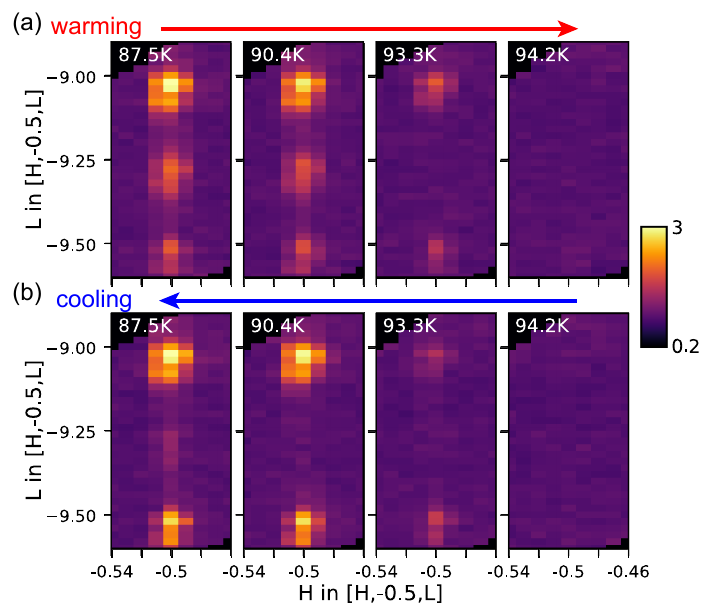}
\caption{Narrow $(H, L)$ maps of reciprocal space of the CDW for a selection of temperatures. (a)(b) show temperature evolution of CDW diffractions in the warming process and cooling process, respectively. The intensity of data in the color map is in logscale.
\label{fig2}}
\end{figure}

To further elucidate the origin of {\bf q}$_1$, {\bf q}$_2$ and {\bf q}$_3$-type CDW peaks, we select the CDW peaks at {\bf q}$_1$ = [-0.5,-0.5,-9], {\bf q}$_2$ = [-0.5,-0.5,-9.5] and {\bf q}$_3$ = [-0.5,-0.5,-9.25] and trace their evolutions across transition temperatures \,\cite{SI}.  As shown in Fig.~\ref{fig2}, the intensities of CDW peaks at {\bf q}$_1$ and {\bf q}$_2$ gradually descend with increasing temperatures and then abruptly drop to zero above $T_{\mathrm{CDW}} \sim$ 94 K. 
Intriguingly, the {\bf q}$_3$-type CDW peak disappears at $\sim$ 93 K, about 1 K lower than that of {\bf q}$_1$ and {\bf q}$_2$-CDWs.
Surprisingly, in the cooling process, while {\bf q}$_1$ and {\bf q}$_2$-type CDW peaks promptly emerge below 94 K, the appearance of {\bf q}$_3$-type CDW peak lags until further cooling below $\sim$ 89K.

\begin{figure}[htbp]
\centering
\includegraphics[width=1\columnwidth]{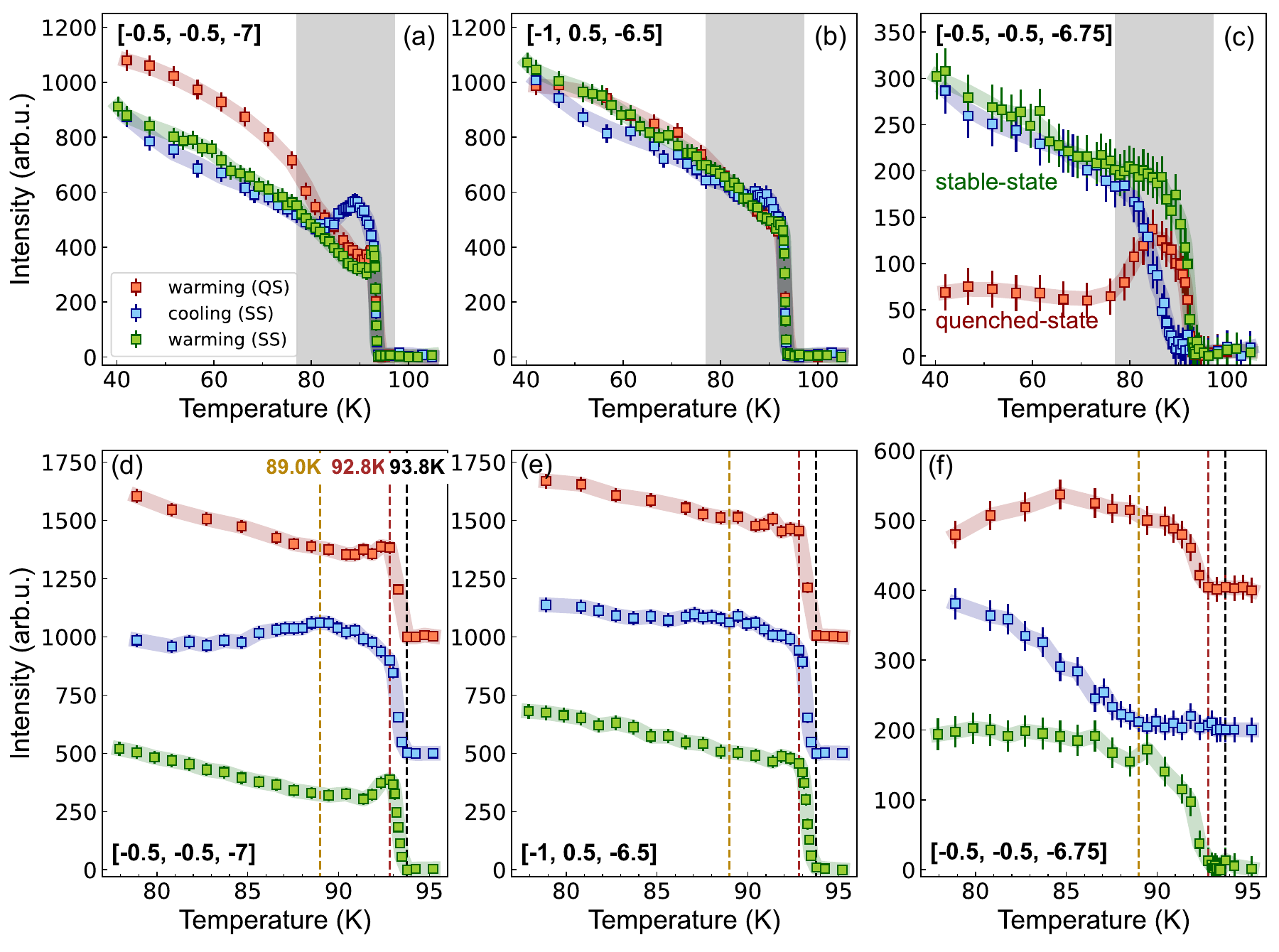}
\caption{(a)-(c) Temperature evolution of three CDW peaks for [-0.5, -0.5, -7], [-1, 0.5, -6.5] and [-0.5, -0.5, -6.75], respectively. The red curve is a warming process after fast cooling, referred as quenched-state (QS) as explained in the text. The blue curve is a cooling measurement and the green curve is another warming process, referred as stable-state (SS) as explained in the text. The error bars are estimated by the square root of counts. (d)-(f) Zoom-in temperature marked by the grey region to visualize three characteristic temperatures, i.e. 93.8 K, 92.8 K and 89 K. The curves are vertically shifted for clarity.
\label{fig3}}
\end{figure}

This thermal hysteresis of {\bf q}$_3$-type CDW peak is unexpected. As further verification, 
the ${\mathrm{CsV}}_3{\mathrm{Sb}}_5$ sample was subjected to multiple thermal cycling, during which the intensities of CDW peaks were recorded \emph{in situ}, as shown in Fig.~\ref{fig3}. The first warming process was measured after a fast cooling from room temperature to $\sim$ 18 K at a rate of $\sim 8$ K/min (red curve). The measurements at each temperature point were repeated six times and summed together for better statistics. The scanning time for each temperature point was around 1.5 hours. After warming up to 105 K, we proceeded to the cool down measurement under the same conditions (blue curve), 
so that CDW was allowed to evolve quasi-statically. Then we performed a second warming measurement (green curve). We have tracked three CDW peaks considering their relatively strong intensity within our measured reciprocal space, indexed by [-0.5, -0.5, 7], [-1, 0.5, -6.5] and [-0.5, -0.5, -6.75] respectively. From these results we have identified a quenched-state (QS) of CDW in ${\mathrm{CsV}}_3{\mathrm{Sb}}_5$ as a consequence of the fast-cooling process, which is evidenced by the discrepancy between the two warming processes (red and green curves). In fact, our experiments revealed that even a freezing rate of 2 K/min was too fast for the 4c CDW to establish completely, and the system would settle in a meta-stable quenched-state with weak $\mathbf{q_3}$-type CDW signals at low temperature \,\cite{SI}. On the other hand, a quasi-static cooling process could drive the 4c CDW into a coherent stable-state (SS) with much high intensity of {\bf q}$_3$-type CDW at low temperature. 

For the first warming process (referred as QS), the {\bf q}$_3$-type CDW peak is greatly enhanced above $\sim$ 75 K as shown in Fig.~\ref{fig3}(c). We note that another XRD study reported the appearance of {\bf q}$_3$-type CDW peak above a similar temperature scale of $\sim$ 60 K upon warming\,\cite{Qstahl2021}. However, instead of the coexistence of {\bf q}$_1$, {\bf q}$_2$ and {\bf q}$_3$-type CDW peaks at 18 K that we observed, they found no signals of {\bf q}$_3$-type CDW peak below 60 K and argued that 4c CDW was a super-cooled phase\,\cite{Qstahl2021}. However, we found that the intensity of the {\bf q}$_3$-type CDW was even stronger after quasi-static cooling, ruling out that possibility. 
Most intriguingly, as shown in Figure~\ref{fig3}(d)-(f), we can identify three characteristic temperatures, i.e. 93.8 K for the onset of {\bf q}$_1$ and {\bf q}$_2$-type CDW peaks, 92.8 K for the onset of {\bf q}$_3$-type CDW peaks during warming process, and 89 K for the onset of {\bf q}$_3$-type CDW peaks during the cooling process. These multiple transition temperatures are contradicting the previous claim that the different modulations observed in CsV$_3$Sb$_5$ would originate from a unique stacking of 2 $\times$ 2 $\times$ 4. The sharp transitions and hysteresis temperature behavior suggest that the CDW transitions are first-order transitions, consistent with other experiments\,\cite{Qstahl2021,J.luo2021NMR,Noah2021prm,NLWang2021prb}.

Moreover, our results suggest a competition between 2c and 4c CDW phases. 
For example, in the cooling process upon the formation of the 4c CDW phase below 89 K, the intensities of CDWs at [-0.5, -0.5, 7] and [-1, 0.5, -6.5] are suppressed after 89 K [see blue curves in Figure~\ref{fig3} and direct comparisons in ref.\,\cite{SI}]. 
That is to say, the onset temperature of 4c CDW at $\sim$ 89 K changes the intensity of {\bf q}$_1$ and {\bf q}$_2$-CDWs from a peak to a dip.
This behavior is more pronounced in the former.
We speculate this is due to the different relative contributions of 2c and 4c CDWs in {\bf q}$_1$ and {\bf q}$_2$-type CDW peaks,
since the structure factors contribute differently in reciprocal space.
A similar competition effect was observed in the warming process, where the disappearance of the 4c CDW phase above 92.8 K led to a peak in the intensity of CDWs at [-0.5, -0.5, 7] and [-1, 0.5, -6.5] [see green curves in Figure~\ref{fig3}]. 

\begin{figure}[htbp]
\centering
\includegraphics[width=0.5\textwidth]{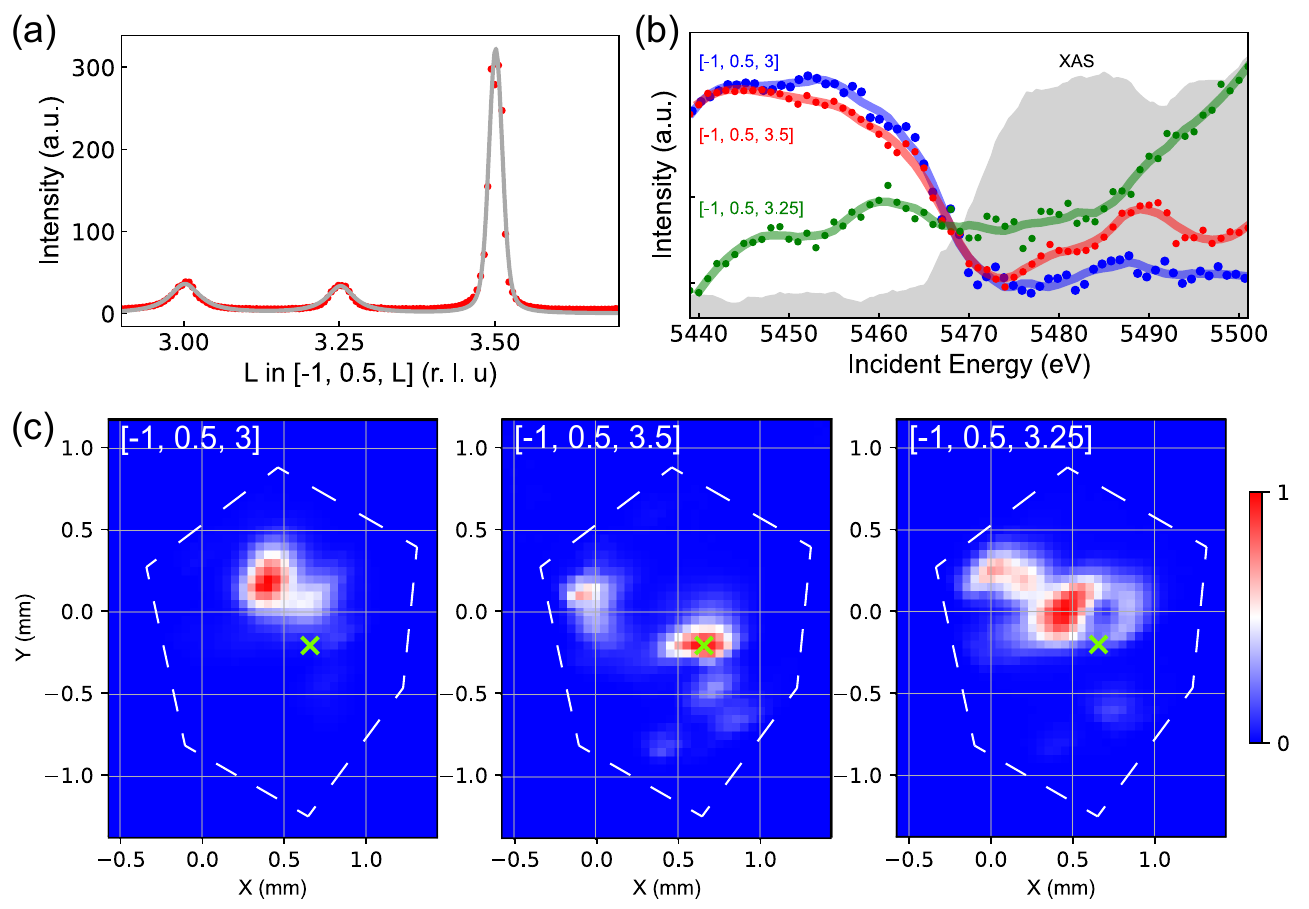}
\caption{(a) $L$-scan of [-1, 0.5, L] measured at 6 K near V $K$-edge (E$_i$ = 5465 eV). The gray curve is the fit to the Voigt function. The measured position is indicated by a cross in panel (c). (b) Energy dependence for three CDW peaks at [-1, 0.5, 3], [-1, 0.5, 3.5] and [-1, 0.5, 3.25], respectively. The solid lines are guides to the eyes. The shaded area indicates the x-ray absorption spectra (XAS) at the V $K$-edge. (c) Spatial distributions for three CDW peaks, respectively. Data were normalized to the respective maximum intensity. The white dashed line indicates the profile of the measured sample. The markers in (c) indicate the sample position measured in (a).
\label{newfig3}}
\end{figure}

The nature of the CDW phase is further elucidated using REXS at V $K$-edge. 
We selected three CDW peaks at {\bf q}$_1$ = [-1, 0.5, 3], {\bf q}$_2$ = [-1, 0.5, 3.5] and {\bf q}$_3$ = [-1, 0.5, 3.25] (see Fig.~\ref{newfig3}(a)) to study their energy dependence. 
As shown in Fig.~\ref{newfig3}(b), we observed a dip at V $K$-edge for all three CDW peaks. This is similar to the energy dependence of the 2 $\times$ 2 $\times$ 2 CDW in ${\mathrm{TiSe}}_2$, which originates from lattice displacement in CDW phase\,\cite{TiSe2_PRR}. 
Then we performed a spatial mapping of the intensity of the three CDW peaks at 5465 eV as shown in Fig.~\ref{newfig3}(c). Intriguingly, the {\bf q}$_1$, {\bf q}$_2$ and {\bf q}$_3$-CDWs seem to exclude each other spatially, rendering a single-phase scenario unlikely.

\begin{figure}[htbp]
\centering
\includegraphics[width=0.5\textwidth]{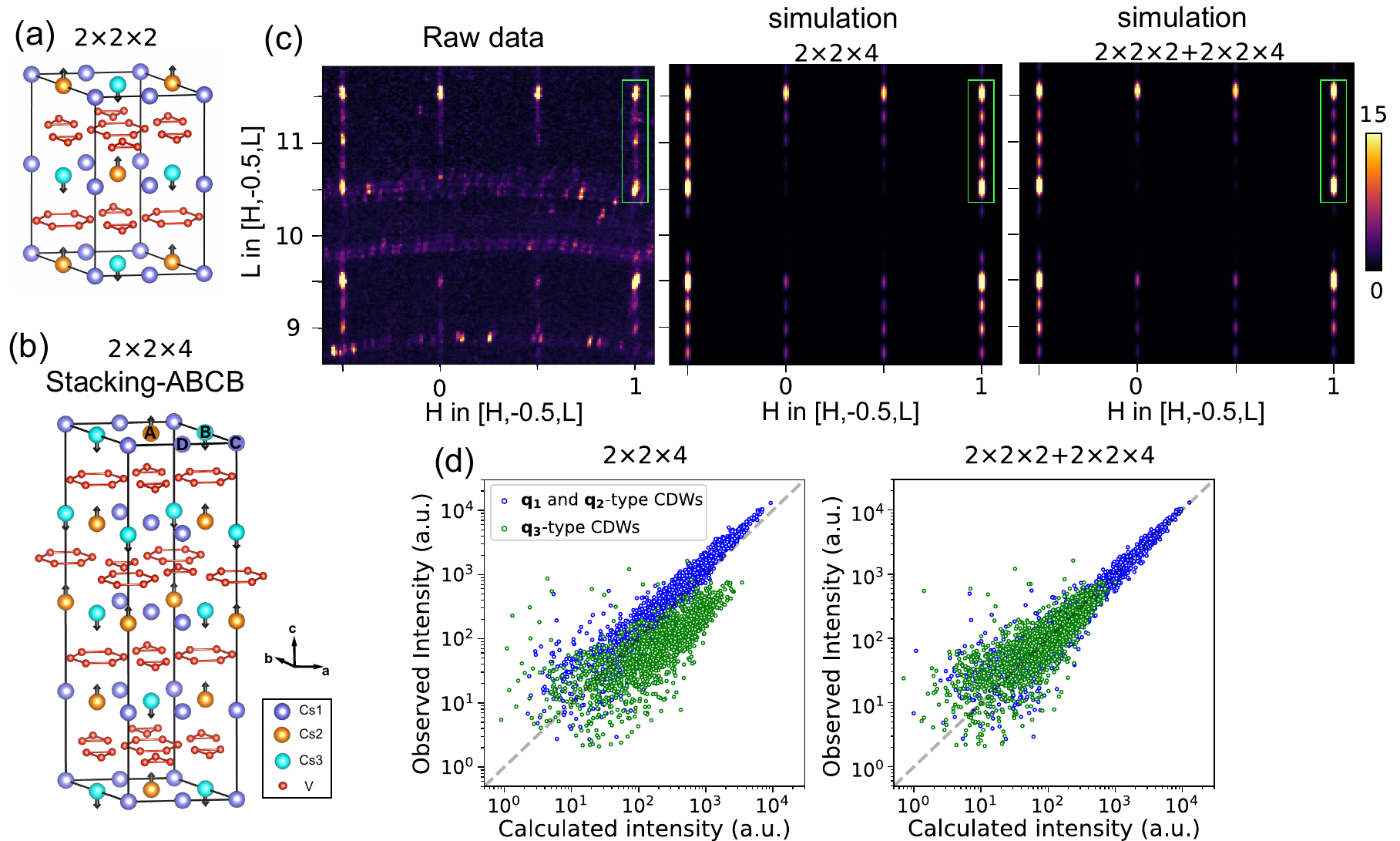}
\caption{(a) and (b) show the 2 $\times$ 2 $\times$ 2 CDW AB-stacking pattern and the 2 $\times$ 2 $\times$ 4 CDW ABCB-stacking pattern, respectively. The comparisons between the experimental XRD pattern and simulation pattern are shown in (c).  The momentum axes are plotted in reciprocal length, i.e., $\AA^{-1}$.
The green rectangle box highlights the difference between the experiment and the two simulations. (d) Observed X-ray diffraction intensities from 3280 CDW peaks compared to our DFT-optimized CDW structures. The blue markers indicate {\bf q}$_1$ and {\bf q}$_2$-type CDW peaks, while the green markers indicate {\bf q}$_3$-type CDW peaks. Considering only 4c CDW structures (left panel), and considering 2c + 4c CDW structures (right panel). 
\label{fig4}}
\end{figure}

Comparison with simulated diffraction patterns lends support for the coexistence of 2c and 4c stacking CDW phases in ${\mathrm{CsV}}_3{\mathrm{Sb}}_5$. 
Our DFT optimization produces three types of 3D CDW structures with approximately the same energy (within 1 meV per formula unit), including one 2c CDW and two 4c CDW \,\cite{SI}.
All structures have identical in-plane ISD deformation and differ in the stackings of V-containing layers. There are 4 translationally inequivalent Cs sites in a $2\times 2$ supercell, denoted A, B, C, D as shown in Figure~\ref{fig4}, and 
an ISD deformed V-layer is called X-stacked (X=A,B,C,D) if its contracting V-hexagon sits on the X-sites. 
In our DFT calculations, the SD deformation is found to be unstable and transforms spontaneously into ISD deformation when neighboring  V-layers are differently stacked \,\cite{H.X.Tan.prl.calcu}. The stable CDW structures are the 2c supercell with AB stacking (see Fig.\ref{fig4}(a)) and the 4c supercells with ABCB (see Fig.\ref{fig4}(b)) and ABCD stacking \,\cite{SI}.
Structural optimization shows that a contracting V-hexagon pulls the neighboring Cs ions toward itself. The frustration of a Cs atom in identifying with the two adjacent V layers then favors alternating stacking of the V hexagons, leading to various possible stackings.

We then fit the experimental XRD results as an incoherent superposition of the intensity of calculated CDW structures. Since $C_6$ symmetry can be broken by the stacking, we consider the $C_6$ equivalents (three for each) in the fittings.
It was argued that all CDW superlattice peaks originated from a single stacking of 4c phase\,\cite{sdh2021Brenden}.
Thus, we first simulated the XRD patterns considering only 4c structures, including three equivalent ABCB sequences and three equivalent ABCD sequences [more information in ref.\,\cite{SI}].
Figure~\ref{fig4}(c) shows representative simulation patterns. The fitting with only the 4c structures is clearly inadequate with exaggerated {\bf q}$_3$-type peaks (correlation coefficient 0.80 for CDW peaks), which can be improved by including both 4c and 2c stackings in the model (correlation coefficient 0.91 for CDW peaks), as is evident from the correlation maps shown in Figure~\ref{fig4}(d). In other words, a reasonable fit of our data is only achieved from the coexistence of 2c and 4c structures. Inclusion of the 2 $\times$ 2 $\times$ 1 (1c) CDW with ISD deformation in the model produces little improvement \,\cite{SI}.
Notably, the proposed 4c CDW structure with modulation between ISD and SD distortions \,\cite{sdh2021Brenden} is inconsistent with our XRD data \,\cite{SI}.

In summary, through systematic temperature-dependent XRD measurements, we discover the coexistence and competition of 2c and 4c CDW phases in ${\mathrm{CsV}}_3{\mathrm{Sb}}_5$ through different transition temperatures. Scanning REXS show that these modulations tend to exclude each other spatially. 
In conjunction with DFT calculation, our diffraction measurements reveal the microscopic origin of 2c and 4c CDW structures arising from different stackings of the ``inverse star of David" structures. These results provide critical insights into the underlying CDW instability in ${\mathrm{CsV}}_3{\mathrm{Sb}}_5$
and offer a possible explanation to understand the double-peak behavior of $T_{\mathrm{c}}$ by applying pressure or doping on ${\mathrm{CsV}}_3{\mathrm{Sb}}_5$\,\cite{F.H.Yu.pressure.nc,ChenKY.pressure.PRL,doping.2sc}. The first peak of $T_{\mathrm{c}}$ may be due to the completed destruction of one stacking phase while the disappearance of both stacking CDW phases leads to the second peak of $T_{\mathrm{c}}$. Our results reveal that the coexistent 2c and 4c CDWs are close in free energy, so external perturbations can effectively modulate the CDW phases to tune the related exotic properties. This establishes ${\mathrm{CsV}}_3{\mathrm{Sb}}_5$ as an exciting new platform for manipulating electronic properties.

\begin{acknowledgments}
Y.~Y.~P. is grateful for financial support from the Ministry of Science and Technology of China (2019YFA0308401 and 2021YFA1401903) and the National Natural Science Foundation of China (11974029). J.F. acknowledges the financial support from the National Natural Science Foundation of China (11725415 and 11934001), the Ministry of Science and Technology of China (2018YFA0305601) and the National Key R\&D Program of China (2021YFA1400100). Y. F. Guo acknowledges the financial support from the National Science Foundation of China (Grant No. 92065201). We acknowledge DESY (Hamburg, Germany), a member of the Helmholtz Association HGF, for the provision of experimental facilities. Parts of this research were carried out at beamline P09 at PETRA III. Beamtime was allocated for proposal I-20220028. We acknowledge Dr.~P.~Bereciartua for helping with the REXS experimental control.
\end{acknowledgments}

\end{document}